**Electron-irradiation-facilitated production of chemically homogenized nanotwins in nanolaminated carbides**

Hui Zhang[1,2]*, Qianqian Jin[3], Tao Hu[4]*, Xiaochun Liu[5], Zezhong Zhang[6], Chunfeng Hu[7], Yanchun Zhou[8], Yu Han[9], Xiaohui Wang[10].

[1]Electron Microscopy Center, South China University of Technology, Guangzhou 510640, China.

[2]School of Emergent Soft Matter, South China University of Technology, Guangzhou 510640, China.

[3]Center for the Structure of Advanced Matter, School of Electronic Engineering, Guangxi University of Science and Technology, Liuzhou 545006, China.

[4]Institute of Materials Science and Devices, College of Materials Science and Engineering, Suzhou University of Science and Technology, Suzhou 215009, China.

[5]Institute of Metals, Changsha University of Science & Technology, Changsha 410004, China.

[6]Electron Microscopy for Materials Research (EMAT), University of Antwerp, Groenenborgerlaan 171, 2020 Antwerp, Belgium.

[7]Key Laboratory of Advanced Technologies of Materials, Ministry of Education, School of Materials Science and Engineering, Southwest Jiaotong University, Chengdu 610031, China.

[8]School of Materials Science and Engineering, Zhengzhou University, Henan 450001, China.

[9]King Abdullah University of Science and Technology, Advanced Membranes and Porous Materials Center, Physical Sciences and Engineering Division, Thuwal 23955-6900, Saudi Arabia.

[10]Shenyang National Laboratory for Materials Science, Institute of Metal Research, Chinese Academy of Sciences, Shenyang 110016, China.

*Corresponding authors. Email: hui.materials.zhang@gmail.com (H.Z.), thu@usts.edu.cn (T.H.)

**Abstract**

Twin boundaries have been exploited to stabilize ultrafine grains and improve the mechanical properties of nanomaterials. The production of twin boundaries and nanotwins is however prohibitively challenging in carbide ceramics. Using scanning transmission electron microscopes as a unique platform for atomic-scale structure engineering, we demonstrate that twin platelets could be produced in carbides by engineering antisite defects. Antisite defects at metal sites in various layered ternary carbides are collectively and controllably generated and the metal elements are homogenized by electron irradiation, which transforms the twin-like lamellae into nanotwin platelets. Accompanying the chemical homogenization, $α$-Ti$_3$AlC$_2$ transforms to unconventional $β$-Ti$_3$AlC$_2$. The chemical homogeneity and the width of the twin platelets can be tuned by the dose and energy of bombarding electrons. Chemically homogenized nanotwins can boost the hardness by ~45%. Our results provide a new way to produce ultrathin (<5 nm) nanotwin platelets in scientifically and technologically important carbide materials and showcase the feasibility of defect engineering by an angstrom-sized electron probe.

**Introduction**

Nanotwinned structures are energetically more stable than normal nanograins with comparable grain sizes as the excess energy of a coherent twin boundary is roughly ten times lower than that of normal grain boundaries[1]. Nanotwinning has thus been used to produce stable ultrafine grains (<5 nm) in metals, e.g. Cu[2] and Ti[3] alloys, and ceramics, e.g. diamonds[4] and cubic BN[5]. The latter generally requires extreme environments such as high pressure and high temperature due to the strong covalent bonds. Interestingly, atomically thin twin-like lamellae naturally exist in ternary layered carbide ceramics, $M_{n+1}AC_n$ phases, which have the merits of both metals and ceramics and hold promise for applications in high-temperature foil bearings, heating elements, electrical contacts, coatings on nuclear materials, etc[6]. In these unique structures, A (A group elements, mainly Al) layers structurally act as mirror planes of the $M_{n+1}C_n$ slabs[7] (Fig. 1a), where M is a transition metal element. The metallic M–A bonding in $M_{n+1}AC_n$ phases is significantly weaker than the ionic and covalent M–C bonding[8]. The twin-like interfaces therefore have no strengthening or hardening effect in $M_{n+1}AC_n$ phases, yet are the Achilles' heel of these structures, where the deformation and delamination occur first[9–11].

It has been expected that the twin-like interfaces could be used to boost the mechanical properties if we can find a way to diminish the chemical anisotropy near the interface. There are two possible routes to achieve this. One may remove A-layers from the structure by chemical etching or heat treatment. The former results in MXenes[12] and weaker structures because of the intercalation of atoms like O, F, H or even molecules. The latter method can produce twins but the process is uncontrollable (Fig. S1). Another way is to chemically homogenize M and A without destroying the mirrored structures. Heat treatment has been used to tune the distribution of constituent elements in twins[13], but is inapplicable for $M_{n+1}AC_n$ phases because of phase decompositions[14]. We thus turned to engineer the antisite defects with the irradiation of energetic particles.

In $M_{n+1}AC_n$ phases, antisite defects $M_A$ and $A_M$ ($M_A$: M at A sites or $A_M$: A at M sites) could form with the bombardment of energetic particles[15,16] with low radiation fluence (~0.2 dpa) due to their relatively low formation energy (2~3 eV for $Ti_3AlC_2$[7]) compared to other point-defect pairs (6~8 eV). The accumulation of antisite defects could lead to the transformation of hexagonal $M_{n+1}AX_n$ to face-centered-cubic $(M_{n+1}A)X_n$ with large amounts of nanotwins[17]. However, due to the large beam size and poor controllability of beam position, the most used heavy ion irradiation is infeasible to spatially engineer the defects at nanometer or smaller scale, which requires atomic-scale atom manipulation. With remarkable improvements in spatial resolution[18] and electron-probe controllability[19], scanning transmission electron microscopes become a new platform to manipulate atoms[20–23], cations[24] and vacancies[25], trigger phase transformations[26] and induce the movement of grain[27] and phase[28] boundaries. Here, we demonstrate that angstrom-sized electron probes can work as an effective stimulator for the solid-state chemical homogenization in various $M_{n+1}AC_n$ phases to produce chemically homogenized twins.

**Results and Discussion**

The three-layer-thick titanium carbide lamellae in $Ti_3AlC_2$ are mirrored with respect to the Al-layers (Fig. 1a,b). Crystallographically, there are two sites for Al at the twin-like plane of titanium carbide lamellae, which leads to two polytypes, i.e. $α$-$Ti_3AlC_2$ (Fig. 1a) and $β$-$Ti_3AlC_2$ (Fig. 1b). The latter has been assumed to exist for decades[29,30] but never conclusively confirmed in experiments before. In the high-angle annular dark-field (HAADF) scanning transmission electron microscopy (STEM) images, the contrast is in proportion to $\overline{Z^n}$ (the average of $Z^n$ for all the atoms in the column along the projection direction), where Z is the atomic number and $n \approx 2$. The atomic columns with high and low intensities in Fig. 1d,g are Ti-layers and Al-layers, respectively. Comparing the atomic positions in Al-layers and Ti-layers (green rectangles) in [$1\bar{2}10$]-images with the structural model (Fig. 1a,b) suggests that $Ti_3AlC_2$ in (d) and (e) takes the $\alpha$- and $\beta$-configurations, respectively. The [$10\bar{1}0$]-images of the two polytypes (Fig. 1g,h) agree well with the model (Fig. 1c). We found that the intensity of Al-layers increases with the accumulation of electron doses along the two zone axes (Fig. 1e,h,f,i, Video S1 and S2), indicating certain amounts of Ti atoms run into Al-layers. As the irradiation dose accumulates to $\sim 4 \times 10^7$ e$^-$/Å$^2$, the intensity of Al-layers reaches ~83% of the intensity of Ti-layers (inset in Fig. 2a). From Fig. 2a, the composition of Ti-layers and Al-layers was estimated to be $Ti_{0.8}Al_{0.2}$ and $Al_{0.4}Ti_{0.6}$, respectively.

It is well-established that HAADF-STEM has a depth of field $\Delta z = \lambda/\alpha^2$ ($\alpha$ is the convergence semi-angle)[31]. The structural features within $\Delta z$ are in-focus and dominate the contrast in the images. The nearly identical intensity of Ti-layers and Al-layers in a region of ~20 nm in thickness imaged with 30 mrad ($\Delta z$=2.2 nm), 20 mrad ($\Delta z$=4.9 nm) and 10 mrad ($\Delta z$=19.6 nm) in Fig. S2 suggests that the repartitioning is not limited to the regions near the surface, but throughout the sample, which is corroborated by the atomic-scale mapping of Ti-K and Al-K signals via energy-dispersive X-ray (EDX) spectroscopy (Fig. 2b,c,d), where the number of Ti atoms in Al-layers gradually increases, while accordingly that of Al atoms in Al-layers decreases to the same level of Ti-layers. As shown in Fig. S3, EDX simulations using quantum excitation of phonon multislice[32] suggest that Al K-edge is more delocalized than that of Ti K-edge, which seemly forms a continuous line instead of atomically resolved individual columns. The core-shell electrons of light elements are weakly bound to their nuclei as compared to heavy elements. As a consequence, light elements exhibit a rather delocalized ionization potential for X-ray generation[33]. But the homogenization of Al is still evident in Fig. 2b-d. Irradiation-induced repartitioning also occurs in other Al-containing MAX phases, e.g. $Ti_2AlC$ (Video S3, S4) and $Nb_4AlC_3$ (Fig. S4). It is noted that there is no noticeable repartitioning in $Nb_2SC$ (Fig. S4), which might be due to the remarkably high formation energy (8.6 eV) of M/A antisite defects compared to Al-containing MAX phases (2~3 eV) like $Ti_3AlC_2$, $Ti_2AlC$, $Nb_2AlC$ and $Nb_4AlC_3$ (Table S1). The repartitioning is readily controllable by electron dose and energy of the impinging electrons (Fig. 2a). The number of Ti atoms redistributed into Al-layers monotonically scales with the dose accumulation. The trends revealed by EDX quantification and semi-quantitative analysis of HAADF-STEM images are pretty consistent, and monitoring the intensity change in HAADF-STEM images directly could give a fairly good estimate of the process in practice.

The elemental process of the repartitioning is the knock-on of Ti, Al and C atoms. Ab-initio molecular dynamics simulation suggests that Ti/Al and C atoms in defect-free $Ti_3AlC_2$ need ~13 eV and ~5 eV to be knocked out (Fig. S5, Video S5, Table S2), which could be sufficiently supplied by the energy transfer from the impinging electrons to $Ti_3AlC_2$ at 300 kV, where Ti, Al and C atoms can be powered with 17.8 eV, 31.6 eV and 70.9 eV, respectively. In real electron-atom interaction, the atoms on the surface are the easiest and earliest to be knocked out, which is known as sputtering[34]. Depending on the surface plane, the energy needed by Ti, Al and C to be sputtered is 2~6 eV (Table S3), which could be easily met by the energy transfer from 200 kV electrons to those atoms, the maximum of which is 11.0 eV for Ti, 19.4 eV for Al and 43.7 eV for C. The sputtering of constituent atoms on the surface creates defects, which will significantly lower the displacement threshold energy of the atoms near the surface and result in a cascade effect. Therefore Ti/Al antisite defects were observed at 200 kV (Fig. 2), despite that the maximum energy Ti could obtain from impinging electrons is slightly lower than the threshold calculated with the defect-free model, 11.0 eV vs. 13.0 eV. Actually, Ti atoms can even be knocked out at 80 kV[23]. The lower energy the incident electrons have, the less energy can be transferred to the atoms within the sample, and the fewer atoms could be powered to the energy threshold needed to move out of their original crystal sites. Therefore, it takes a higher electron dose at 200 kV (black spheres in Fig. 2a) than 300 kV (green spheres in Fig. 2a) to achieve the same level of atomic repartition. At 200 kV, Ti atoms in Al-layers approach the limit of 75% with a dose of ~$4.7 \times 10^8$ e$^-$/Å$^2$, where nearly complete homogenization is attained. With the complete homogenization of Ti and Al, the structure becomes nanotwins. The twin platelets are equivalent to the $\Sigma 3\{111\}$ twin lamellae in a face-centered cubic structure. With further irradiation, the adjacent twin platelets could be merged into a thicker one, e.g. ~3 nm in Fig. S6.

Carbon atoms are much easier to be knocked out considering the low displacement threshold energy (~5 eV) and high kinetic energy (~71 eV at 300 kV) they can get from the impinging electrons. But we were unable to see where the displaced C atoms stay in HAADF-STEM images (Fig. 1) due to the well-known intrinsic limitation of this imaging modality. Alternatively, we turned to four-dimensional (4D) STEM ptychographic phase imaging, which has remarkable light-element sensitivity[35]. C columns are unequivocally resolved in the phase image (black arrows in Fig. 3a). The remarkably inhomogeneous intensities of the C-layers marked by black solid arrows suggest that some displaced C atoms stay at the interstitial sites of carbide lamellae.

More importantly, some displaced C atoms move to the space between Al-layers and Ti-layers (black hollow arrows in Fig. 3a). Interestingly, the homogenization of C is significantly slower than that of Ti and Al, which might be due to the highly movable nature of C (Video S5). The time resolution of the current microscope is at the scale of microseconds, which is too long to be able to reveal the dynamical repartitioning process and shed insights on the counterintuitive repartitioning behavior of C.

Accompanying the repartitioning of constituent atoms, $α$-polytype (Fig. 1d,g) transforms into $β$-polytype (Fig. 1e,h,f,i). We observed that $α$- and $β$-polytypes coexist between nanotwins and their pristine counterparts, exhibiting line-like contrasts in the HAADF-STEM images (Fig. 3b). The line profile of the line-like layer (transition region in Fig. S7) shows two different periodicities, $d_1$=0.28 and $d_2$=0.1 nm (Fig. 3c). The former corresponds to the distance between the Al columns in $α$- or $β$-polytype, while the latter is the space of Al columns between $α$- and $β$-polytype (Fig. 3d). The pairs with 0.1 nm distance are clearly resolved in the ptychographic phase image (green arrows in Fig. 3a).

The $β$-polytype is metastable and has never been synthesized in experiments before[7,36]. However, it gradually becomes thermodynamically more favorable than the most common $α$-polytype with the formation of $Ti_{Al}$ and $Al_{Ti}$ antisite defects (Fig. 4a). Specifically, the value of $E_β$-$E_α$ becomes negative when 32~48% (~40% on average) Al in Al-layers are replaced by Ti, suggesting $α$-to-$β$ transition may occur if 32~48% $Ti_{Al}$ forms in Al-layers, which is remarkably consistent with the experiments where we observed that $β$-polytype forms when ~30% Al was replaced by Ti (Fig. 1e). To elucidate the stabilization mechanism of $β$-polytype, we closely inspected the electron structures via partial density of states (pDOS, Fig. 4b). In the energy range from -5 eV to the Fermi energy ($E_F$), Ti 3d electrons are dominant compared to Al and C. For $α$- polytype, the density of states of antisite (aTi) and neighboring Ti (sTi) at $E_F$ are 1.47 a.u. and 0.46 a.u., respectively. The significant difference in the density of states between aTi and sTi suggests that breaking Ti–C bonds may render considerable amounts of 3d electrons in $\alpha$-polytype unsaturated. While in $\beta$-polytype, the density of states of aTi and sTi at $E_F$ are close (0.89 a.u vs. 0.88 a.u) and the peaks at -0.40~-0.10 eV are largely overlapped, indicating that orbital hybridization occurs between aTi and sTi, which reduces the unsaturated 3d electrons. We thus speculated that the pronounced difference in electronic structures dictates the relative stability of $\alpha$- and $\beta$-polytype. To gain more insights into this issue, the bonding states of aTi, aAl, sTi and C atoms were calculated via the crystal orbital Hamilton population (COHP) curves[37]. As can be seen in Fig. 4c, nonbonding states of aTi prevail in the highest occupied bands close to $E_F$ in $\alpha$-polytype, while $\beta$-polytype is massively characterized by aTi-sTi and aTi-aAl bonding interactions. Integrated COHP of $\beta$ is more negative than that of $\alpha$ for orbitals (Table S4), confirming stronger bonding interactions in $\beta$, which makes $\beta$-polytype more stable.

Angstrom-sized electron probes can serve not only as a markedly coherent "light source" for advanced real-space imaging but also as a highly controllable power source for atoms to jump out of their potential valley. Using the state-of-the-art transmission electron microscope, we homogenized M and Al elements in various $M_{n+1}AlC_n$ and produced ultrathin (<5 nm) carbide nanotwin platelets. The homogenization process is readily controllable by electron dose and energy. Accompanying the homogenization, the unconventional $\beta$-polytype forms in $Ti_3AlC_2$. More importantly, the irradiated region could dilate ~3.0% along the *c*-axis (Fig. S15), contract by 0.6% along the *a*-axis and expand by ~1.9% in volume. The slight volumetric expansion exerts compressive stress on the untransformed region. In addition, chemical homogenization removes the bonding anisotropy of $Ti_3AlC_2$ and turns the weak twin-like interface into strong twin boundaries of $(Ti,Al)_3C_2$. We speculate that these two factors have positive effects on mechanical properties. As a proof-of-concept study, we irradiated $Ti_3AlC_2$ bulk sample (Fig. S16) and measured the micro-hardness. Ti and Al were nearly completely homogenized in the ~3 nm subsurface region (Fig. S17), and the microhardness was improved by ~45% (Fig. S18).

**Summary**

We demonstrate that nanotwins (<5 nm) could be produced in various $M_{n+1}AC_n$ phases (e.g. $Ti_2AlC$, $Ti_3AlC_2$ and $Nb_4AlC_3$) by electron irradiation and the width of twin platelets is tunable by manipulating electron doses. In the most extensively investigated $M_{n+1}AC_n$ phase, $Ti_3AlC_2$, electron irradiation produces massive metal antisite defects in a readily controllable way. The accumulation of those antisite defects chemically homogenizes the twin-like structure and turns the twin-like interface into twin boundaries. Using atomic-resolution ptychography, the carbon atoms were observed in carbides for the first time, which provides direct evidence of the homogenization of carbon atoms as electrons bombard $Ti_3AlC_2$. With the formation of nanotwinned platelets, the micro-hardness increases by ~45%. In addition, we unequivocally prove that $Ti_3AlC_2$ can transform from $\alpha$- to $\beta$-polytype with the irradiation of electrons. The transformation is also corroborated by first-principle calculations. The universality and tunability of chemical repartitioning by electron irradiation provide a feasible tool for defect and crystal-structure engineering and open a huge space to produce high-performance nanotwinned transition metal carbides.

**Experimental Section.**

*Sample preparation and TEM observation.* Four MAX phases were used in this study, i.e. $Nb_2SC$, $Ti_2AlC$, $Ti_3AlC_2$ and $Nb_4AlC_3$. They were synthesized by reactive hot processing as reported in Refs. 38–40. For (scanning) transmission electron microscopy ((S)TEM) observations, electron transparent specimens were prepared by ion milling. Most (S)TEM characterizations were performed on a Themis Z microscope equipped with double aberration correctors. The convergence semi-angle used in the study was 21 mrad. The HAADF-STEM images were recorded with a collection semi-angle of 65 mrad. The atomic-scale elemental mapping was performed on JEM-ARM200F with a 0.96 sr EDX spectrometer. To estimate the electron dose, the probe current was calibrated with the Faraday cup. The electron dose was calculated by $D = 6.242 \frac{x \times t}{\Delta d \times \Delta d}$, where $D$, $x$, $t$ and $\Delta d$ are dose in $e^-/Å^2$, probe current in pA, dwell time on each pixel in μs and pixel size in Å, respectively. 4D-STEM datasets were recorded on EMPAD[41]. The camera length was chosen to ensure that the maximum collection semi-angle was roughly three times of the convergence semi-angle (21 mrad). Given the long dwell time on each pixel, 1 ms, a probe current of ~5 pA was used to avoid sample damage. The ptychographic phase images were reconstructed by the single-side band method[42,43].

*STEM simulation.* HAADF-STEM simulations were performed with QSTEM[44], where the model was set to be ~25-nm-thick based on the thickness measurement by electron energy loss spectrum. The models were sliced into 1-Å-thick layers. The frozen phonon method was used to address the thermal diffuse scattering, where the results were averaged over 30 frozen phonon configurations. The residual aberrations measured after tuning the corrector and an inner collection semi-angle of 65 mrad were set in the simulation. The outer detector angle was 200 mrad. The Debye-Waller factor of Ti, Al, and C were 0.44, 0.075 and 0.81Å, respectively.

To semi-quantitatively determine the amount of Ti atoms repartitioned to Al-layers in $Ti_3AlC_2$, a library of simulated HAADF-STEM images was established. With the electron irradiation, Ti atoms in Ti-layers move into Al-layers and the corresponding amount of Al atoms in Al-layers go into Ti-layers accordingly. As demonstrated in the ptychographic phase image, carbon can migrate to the interstitial sites between Ti-layers and Al-layers of the pristine structure. However, carbon atoms do not affect the intensity analysis of HAADF-STEM images due to their low atomic number compared to Ti and Al, so the redistribution of carbon atoms was not considered in the image simulation. Five models with $x$ = 0, 0.05, 0.1, 0.15, 0.2, 0.25 for $(Ti_{1-x}Al_x)_3(Al_{1-3x}Ti_{3x})C_2$ were used to establish the lookup table. The intensity ratio between Al-layers and Ti-layers, $I_{Al}/I_{Ti}$, was calculated from the simulated images and fitted with $x$. After that, $I_{Al}/I_{Ti}$ ratios for the experimental images were also calculated and averaged among the values obtained from various atomic layers in the same irradiated region to reduce experimental errors. The concentrations of Al in Ti-layers and Ti in Al-layers were finally calculated using the relationship ($x-I_{Ti}/I_{Al}$) established with the simulated images.

*DFT calculation*. Density functional theory (DFT) calculations were performed to model the electron beam interaction with $Ti_3AlC_2$. The calculations were performed with the projector-augmented wave method using Vienna *Ab initio* Simulation Package[45]. GGA-PBE type exchange-correlation functional was used for all calculations[46]. The equilibrium structure of $Ti_3AlC_2$ was obtained by structural optimization. The total energies and Hellmann-Feynman forces acting on atoms were converged to $10^{-6}$ eV and 0.005 eV/Å$^{-1}$, respectively. Monkhorst-Pack scheme k-point meshes smaller than 0.5 Å$^{-1}$ were used for the integration in the irreducible Brillouin zone. After geometric optimization, *ab initio* molecular dynamics was performed in the NVT ensemble to obtain the displacement threshold energy ($E_d$) and sputtering energy ($E_s$). An orthogonal supercell containing 96 atoms was constructed from the hexagonal unit cell, ensuring the simulation box is larger than 10 Å along the ***a-***, ***b-*** and ***c***-axis to avoid image interactions. $E_d$ was determined by running a series of *ab initio* molecular dynamics simulations with different initial kinetic energies. We used a time step of 1 fs and the maximum time duration up to ~2 ps to ensure that the system converges to equilibrium states. The initial kinetic energy added to the primary knock-on atom (KPA)[47] along [0001] was increased with a step of 1 eV until the atom moved out of their original sites. The KPA not returned to its original position within 2 ps was regarded to be a displaced atom. $E_d$ is the initial kinetic energy to displace KPA. As atoms on the surface are bonded differently from the atoms within the bulk, $E_s$ of atoms on the surface were also calculated. Three typical surfaces, e.g. ($11\bar{2}0$), ($10\bar{1}0$) and (0001), were modeled, as shown in Fig. S19. The slab models used in the study were at least 8-atom-layer thick and no obvious surface reconstructions were observed in the optimized models without any initial kinetic energies added to the constituent atoms. After the structural optimization, the initial kinetic momentum was added along [0001], and $E_s$ was determined in a way similar to $E_d$.

**Supporting Information**

TEM images of annealed $Nb_4AlC_3$ and $Ti_3AlC_2$ with various convergence angles, simulated EDX of $Ti_3AlC_2$, irradiation effect of $Nb_4AlC_3$ and $Nb_2SC$, *Ab initio* molecular dynamics of $Ti_3AlC_2$, HRTEM images of nanotwins with various thicknesses and transition regions, relaxed structures of $Ti_3AlC_2$ with various concentration of antisite defects, hardness of the irradiated sample.

**Acknowledgments:** We thank the National Center for Electron Microscopy in Lawrence Berkeley National Laboratory and Monash Center for Electron Microscopy for microscope access during the initial stage of this project.

**Figures**

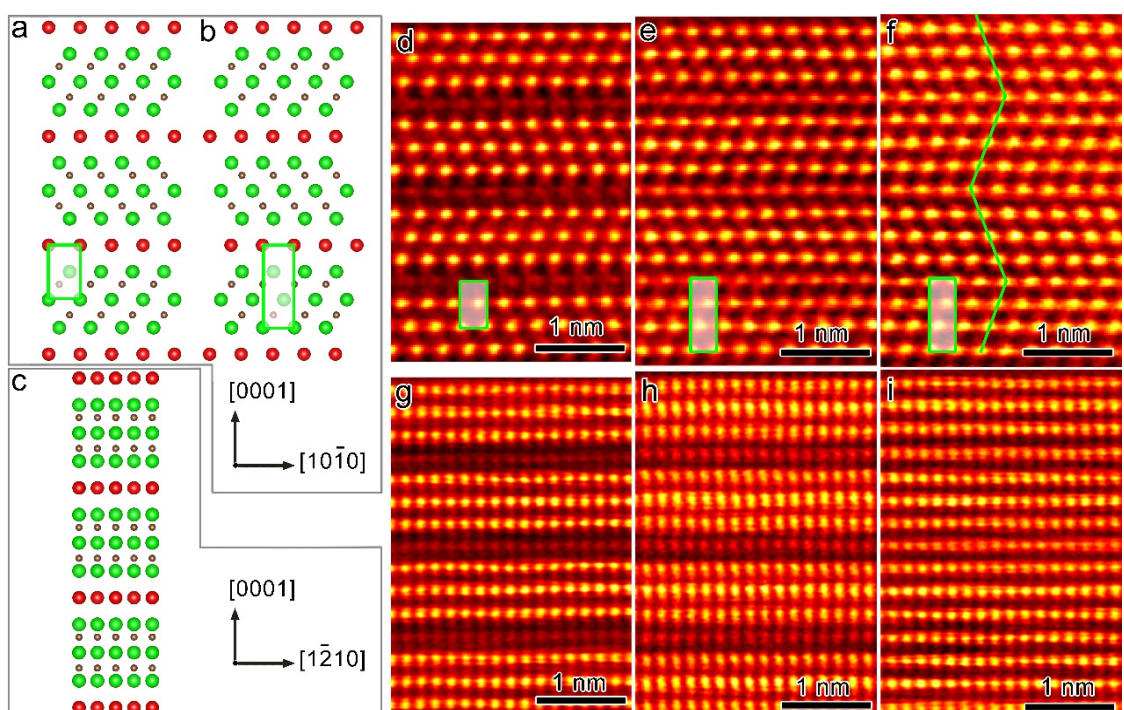

Fig. 1. Electron-irradiation-facilitated generation of carbide nanotwins. Projection of (a) $\alpha$-Ti$_3$AlC$_2$ and (b) $\beta$-Ti$_3$AlC$_2$ atomic models along (a,b) [1$\bar{2}$10] and (c) [10$\bar{1}$0]. [10$\bar{1}$0]-projections of the two polytypes are the same. (d,g) HAADF-STEM images of pristine $\alpha$-Ti$_3$AlC$_2$. Ti, Al and C atoms are denoted by green, red and dark-yellow balls, respectively. In (e,h) and (f,i), the structures have been irradiated by ~2 × 10$^7$ and ~4 × 10$^7$ electrons per square angstrom, respectively. Green lines in (f) illustrate the planes corresponding to {111} of the face-centered cubic structure. The twin platelets are ~1 nm thick. The rectangles in (a,b) and (d–f) highlight the atoms in Ti-layers that are on the same vertical line with the atoms in Al-layers. The structure in (d) and (e,f) is $\alpha$- and $\beta$-polytypes, respectively.

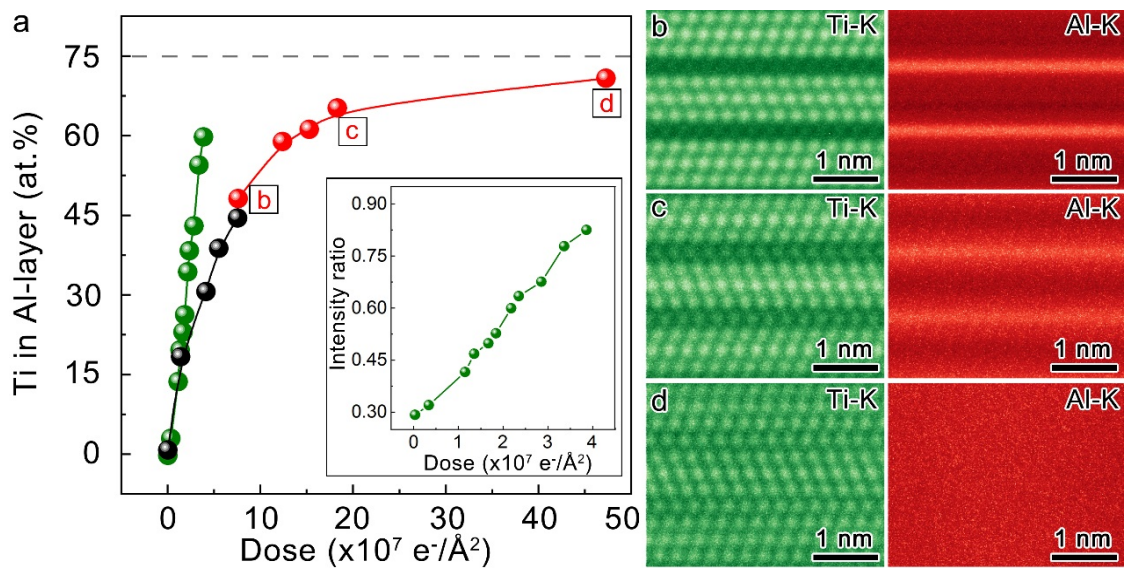

Fig. 2. Irradiation-facilitated element repartitioning in $Ti_3AlC_2$. (a) The concentration of Ti (*at*.%) in Al-layers versus electron doses. The green and black spheres are the data for 300 kV and 200 kV, respectively. The red spheres are EDX data collected at 200 kV. Dash line denotes the concentration limit where Ti and Al are completely homogenized. The concentrations were determined by semi-quantitative analysis of HAADF-STEM images (green and black spheres) and EDX quantification (red spheres). Inset shows the intensity ratio of Al-layers to Ti-layers in the structure irradiated by various doses. (b,c,d) Atomic-resolution EDX maps corresponding to the points marked in (a).

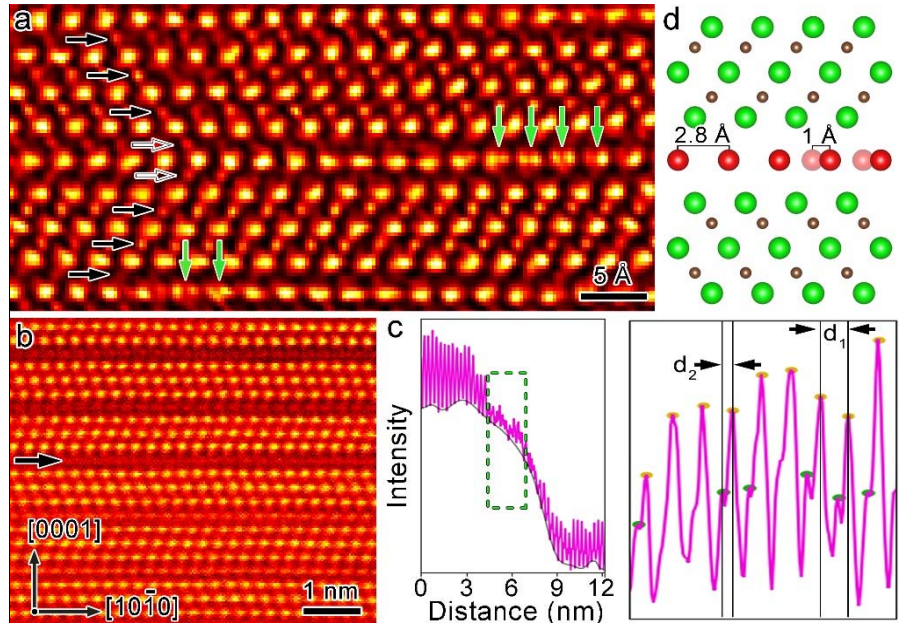

Fig. 3. Structural transition region in irradiated $Ti_3AlC_2$. (a) Ptychographic phase image of the formed twins. The sample was irradiated by ~4 × $10^7$ $e^-/Å^2$. Solid and hollow black arrows mark carbon columns. Green arrows highlight the columns where the atomic positions of Al layers in $α$ and $β$ are both occupied. (b) Line-like layer, marked by the black arrow, near the transformed $Ti_3AlC_2$ (lower half). Nanotwins formed in the bottom region. (c) Intensity profile along line-like layers between the pristine and transformed regions in Fig. S7. The green dash rectangle region is enlarged on the right. $d_1$ = 0.28 nm, $d_2$ = 0.1 nm. (d) Projection showing Al columns in $α$ (red balls) and $β$ (shadowed red balls) polytypes. Ti and C columns are denoted by green and dark-yellow balls, respectively.

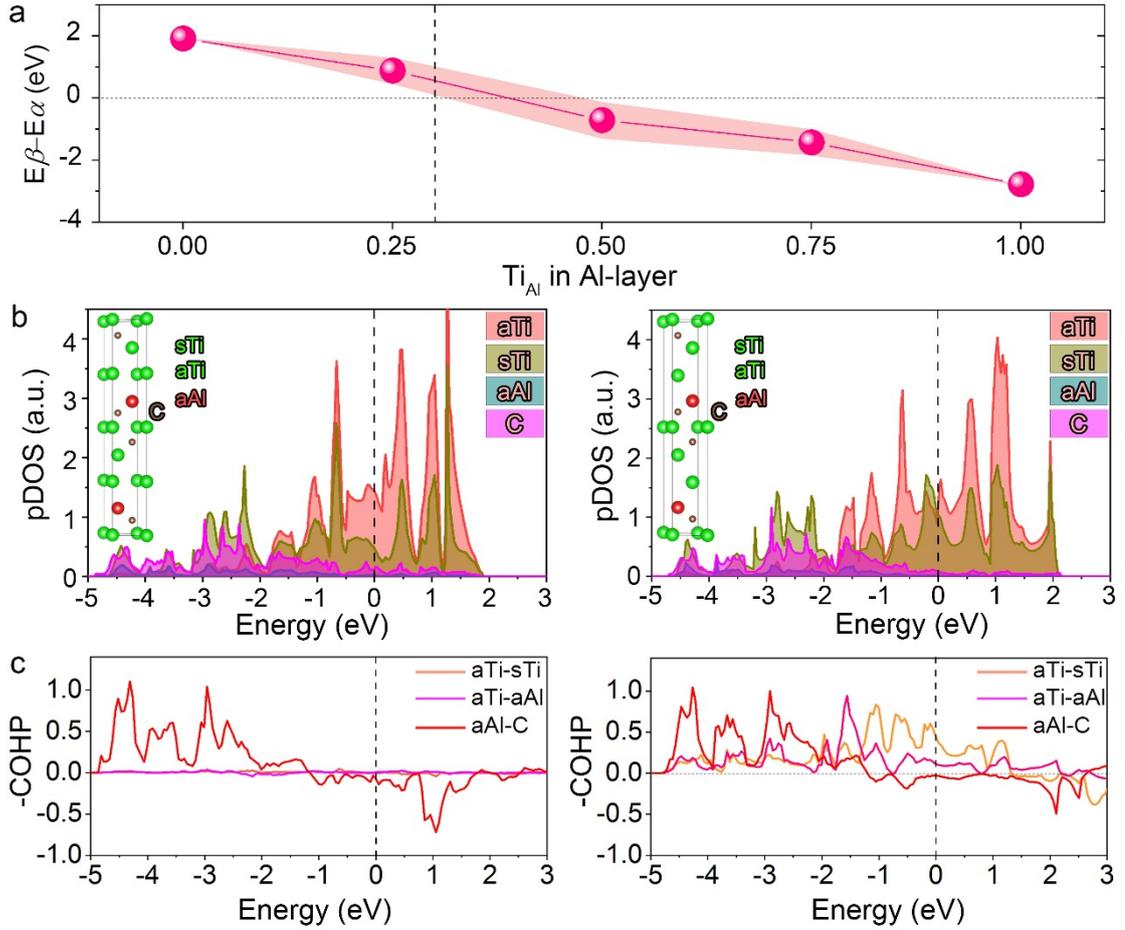

Fig. 4. Energetics and bonding analysis of polytypic transformation. (a) Energy difference ($E_\beta$-$E_\alpha$) between $\beta$ and $\alpha$ with 0, 25%, 50%, 75% and 100% Al in Al-layers replaced by Ti atoms near Al-layers. The shadowed region denotes the difference range obtained from the calculations with multiple configurations shown in Fig. S8-S13. Red spheres are the averaged values. Negative values mean that $\beta$ is energetically favorable. Dash line marks the ratio of Al replaced by Ti in Fig. 1e. (b) pDOS of aTi 3d, sTi 3d, aAl 3p and C 2p orbitals in $\alpha$ (left panel) and $\beta$ (right panel) with Al atoms in Al-layers completely replaced by Ti atoms just below Al-layers. Ti atoms in Al-layers are denoted by aTi, while those above Al-layers are indicated by sTi. Al atoms in Ti-layers are referred to as aAl. Insets show the positions of aTi, sTi, aAl and C. (c) COHP curves of $\alpha$ (left panel) and $\beta$ (right panel) for Ti-Al antisite pairs (aTi and aAl) and their neighboring atoms (sTi and C). $E_F$ was set to zero. Positive, zero and negative -COHP

values signify bonding, nonbonding antibonding interactions, respectively. The models used for (b,c) was displayed in Fig. S14.

**Supporting Information**

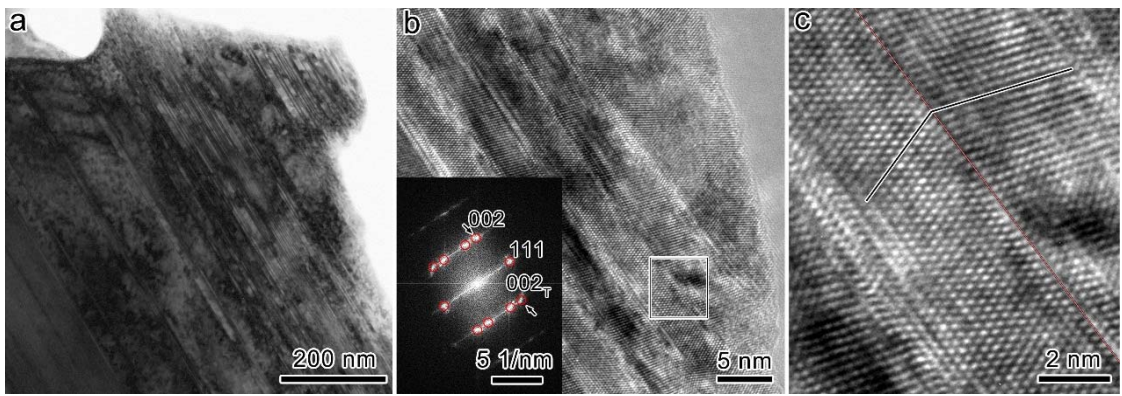

Fig. S1. Nanotwins in annealed Nb$_4$AlC$_3$ MAX phase. (a) Low-magnification transmission electron microscopy (TEM) image of nanotwins. (b) High-resolution TEM image of the twin platelets in (a). Inset is the fast Fourier transform pattern. (002) and (002$_T$) are the diffraction spots of the parent and twinned face-centered cubic crystals. The squared region is further enlarged in (c), where the black and red lines denote the twinned {111} planes and twin boundary, respectively. Nb$_4$AlC$_3$ was annealed at 1900°C for 12 h.

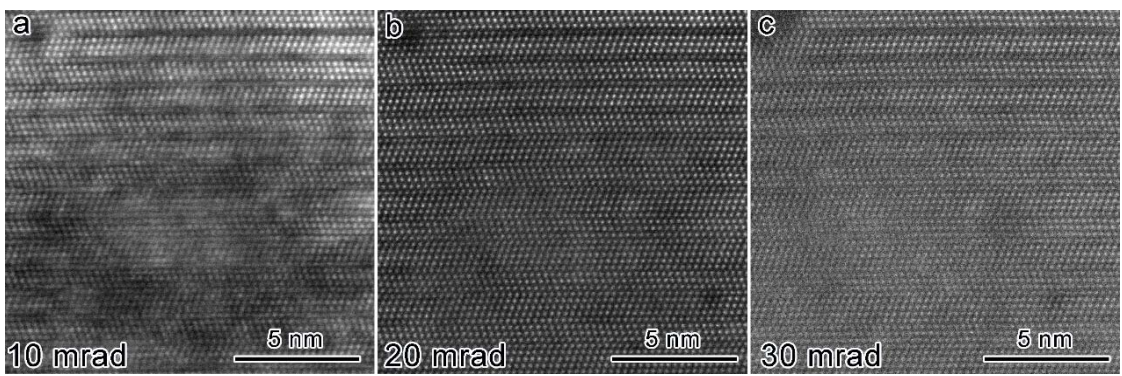

Fig. S2. HAADF-STEM image of irradiated Ti$_3$AlC$_2$ at 300 kV. The irradiated region was imaged with a convergence semi-angle of (a) 10 mrad, (b) 20 mrad and (c) 30 mrad. The collection semi-angle in (a), (b) and (c) was ~30 mrad, ~60 mrad and ~90 mrad, respectively. Ti-layers and Al-layers of Ti$_3$AlC$_2$ in the transformed region have nearly the same level of intensity, regardless of convergence semi-angle.

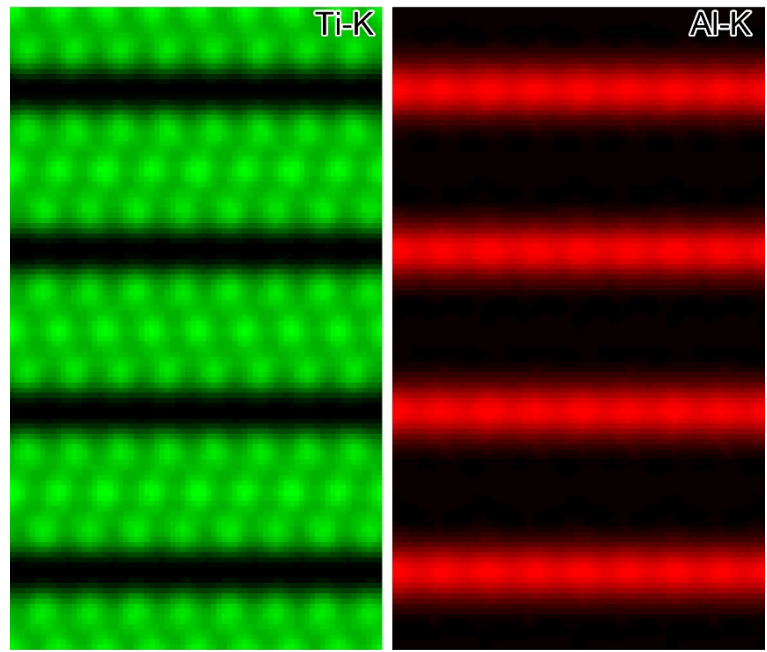

Fig. S3 Simulated EDX maps of Ti (left) and Al (right). The EDX simulations were performed with muSTEM. We used a pixel size of 0.03 Å, a probe step size of 0.3 Å and 30 phonon configurations with experimental microscope parameters.

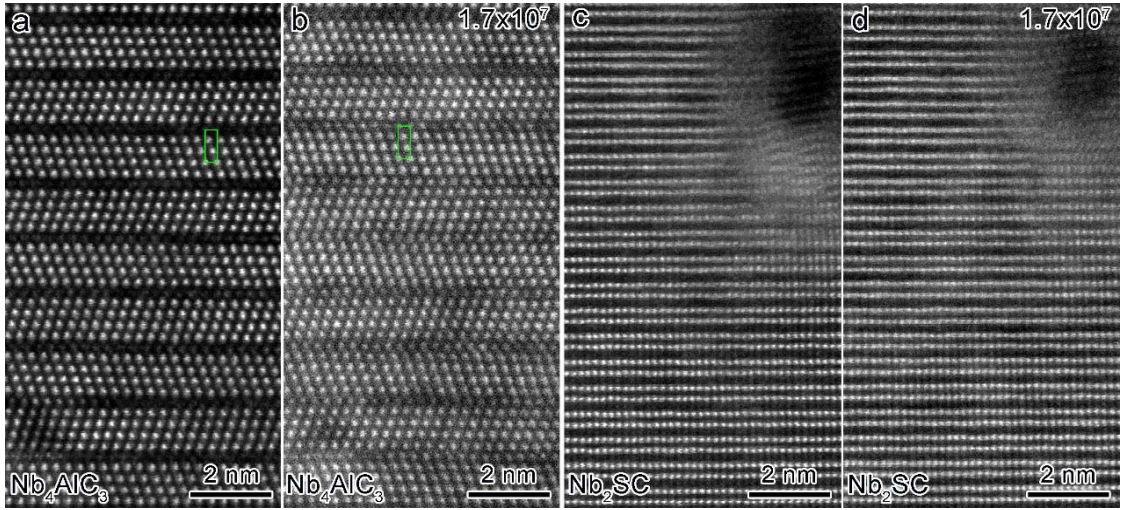

Fig. S4. Irradiation effect of A element in MAX phases. HAADF-STEM images of (a,b) $Nb_4AlC_3$ and (c,d) $Nb_2SC$. (a) and (c) show the pristine structure. The bright layers are Nb layers, and those between Nb layers are Al or S layers. (b) and (d) are the structures irradiated with a dose of ~$1.7×10^7$ $e^-/Å^2$. Certain amounts of Nb have been repartitioned into Al-layers during sample tilt and microscope alignment. After being irradiated with a dose of ~$1.7×10^7$ $e^-/Å^2$, Al-layers have nearly the same level of intensity as Nb-layers. While the intensity of S-layers in (d) is almost unchanged compared to that in (c). Green rectangles in (a,b) mark the Nb atoms in carbide lamella that are on the same veritical line with the atoms in Al-layers. No polytypic transformation was observed in $Nb_4AlC_3$.

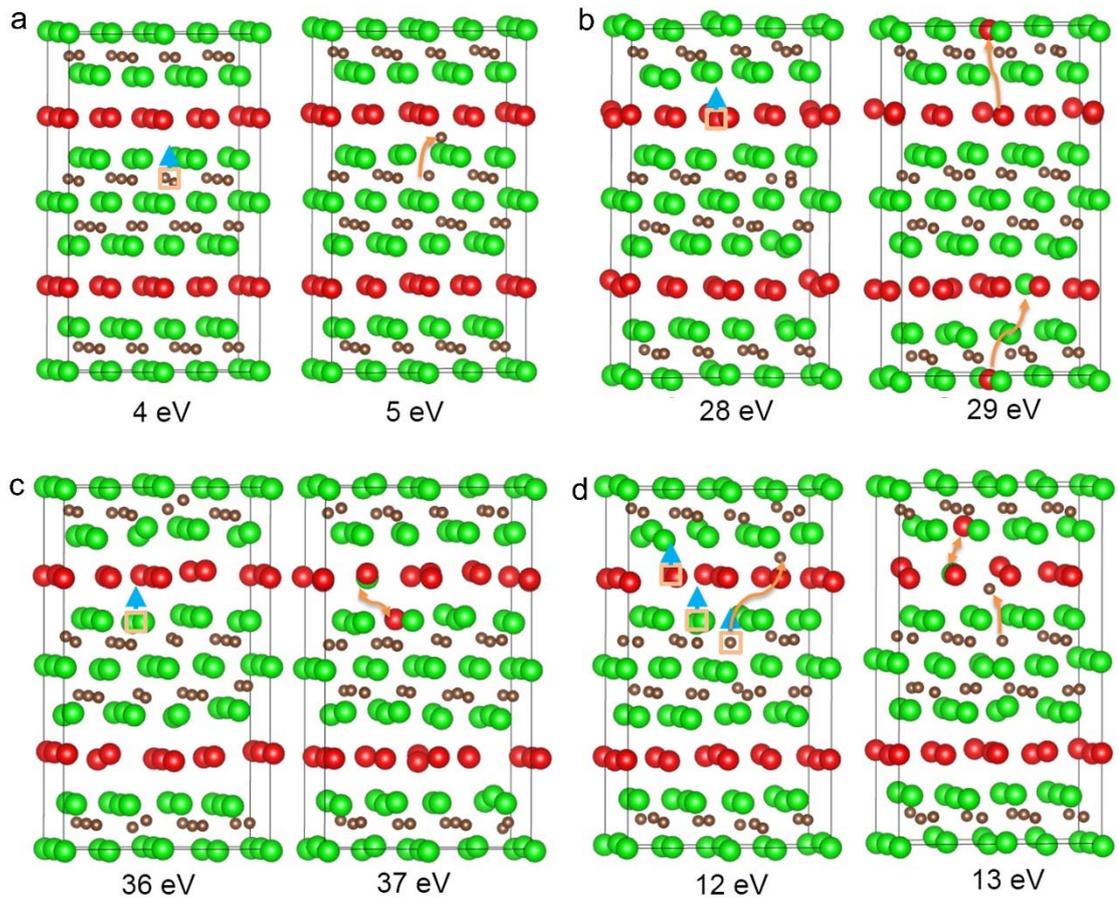

Fig. S5. *Ab initio* molecular dynamics of $\alpha$-Ti$_3$AlC$_2$. Atomic configurations after 2 ps equilibration with C (a), Al (b) and Ti (c) atoms added an initial kinetic energy of 4~5 eV, 28~29 eV and 36~37 eV, respectively. The energy is presented at the bottom. In (d), Ti, Al and C were all powered with the same kinetic energies. Blue arrows denote the direction of the initial momentum. C interstitials and Ti/Al antisite defects are marked by orange arrows. Ti, Al and C atoms are denoted by green, red and yellow balls, respectively.

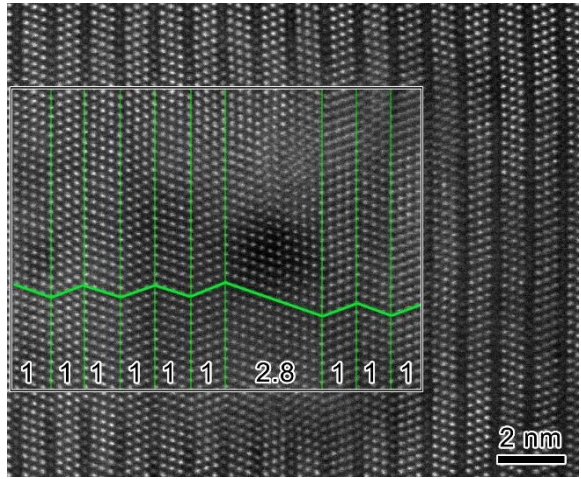

Fig. S6. Nanotwins with various thicknesses. The outlined region was irradiated with ~5×10$^8$ e$^-$/Å$^2$ at 200 kV. The twinned planes and twin boundaries are highlighted by green lines. The width in nanometers is denoted at the bottom.

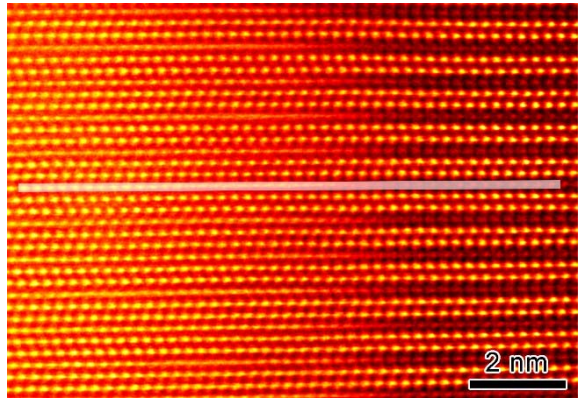

Fig. S7. Transition region between pristine and transformed structures. HAADF-STEM image shows line-like layers between the pristine (right) and transformed (left) regions. The intensity profile in Fig. 3c was extracted from the line position.

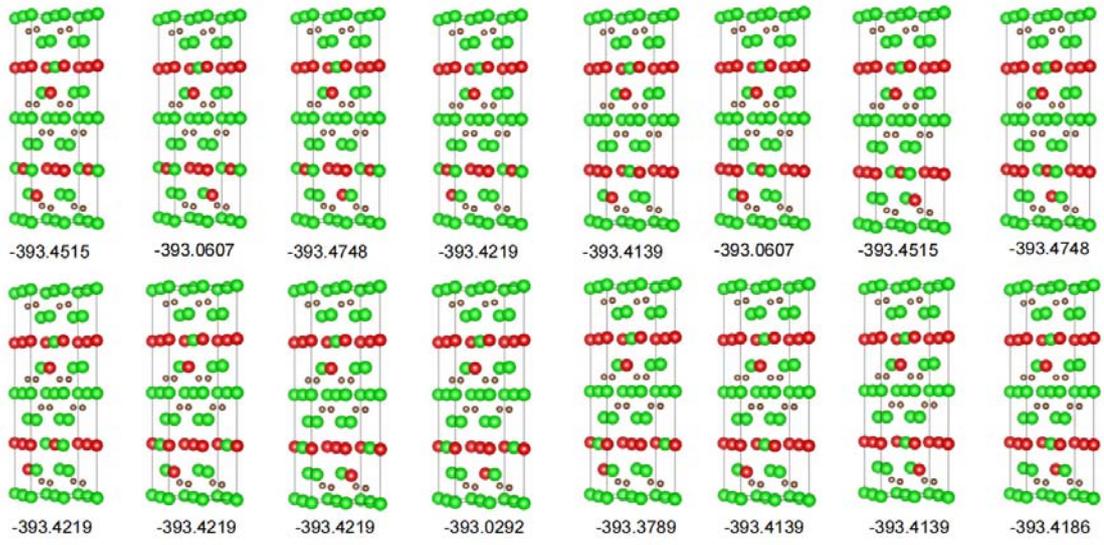

Fig. S8. Relaxed structures of α-Ti$_3$AlC$_2$ with 25% Al atoms forming Ti-Al antisite defects. The total energies in eV of stable configurations are denoted at the bottom. Ti, Al and C atoms are denoted by green, red and yellow balls, respectively.

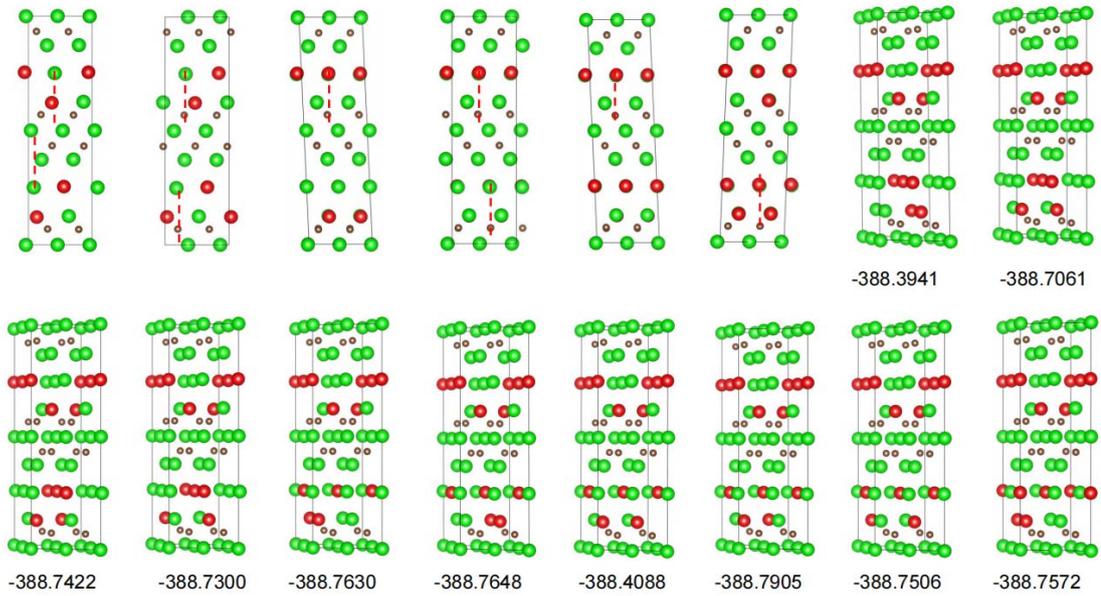

Fig. S9. Relaxed structures of $\alpha$-Ti$_3$AlC$_2$ with 50% Al atoms forming Ti-Al antisite defects. The total energies in eV of stable configurations are denoted at the bottom. There are six structures without total energies shown here, where the atoms in Al-layers are on the same vertical line with the atoms in the near Ti-layer (the first) or C-layer (the second to the sixth). $\alpha$-configuration in those structures is not stable. Ti-Al antisite defects in the second to sixth structures induce partial transformation of $\alpha$ to $\beta$. Ti, Al and C atoms are denoted by green, red and yellow balls, respectively.

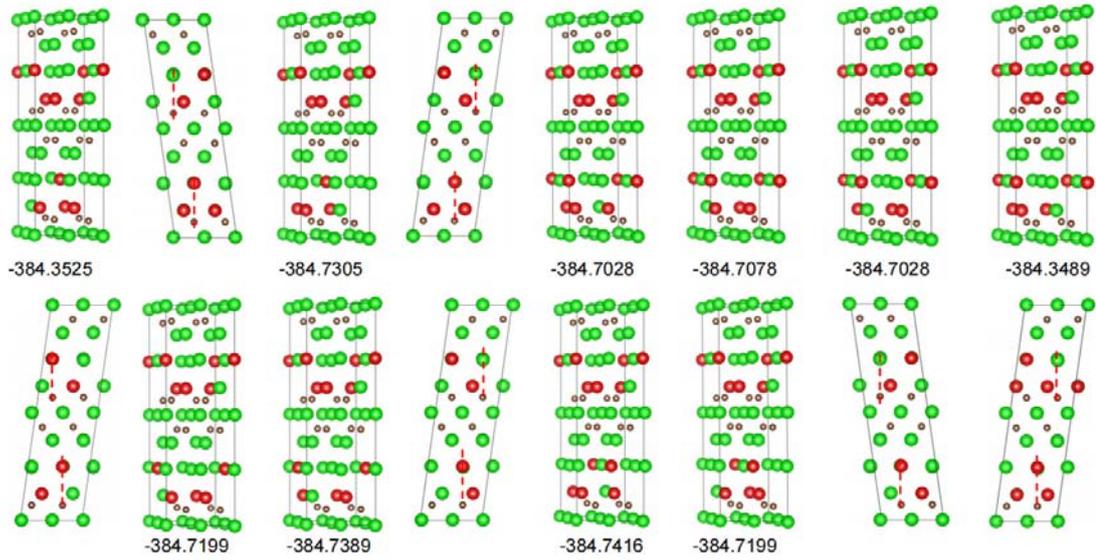

Fig. S10. Relaxed structures of $\alpha$-Ti$_3$AlC$_2$ with 75% Al atoms forming Ti-Al antisite defects. The total energies in eV of stable configurations are denoted at the bottom. There are six structures without total energies shown here, where the atoms in Al-layers are on the same vertical line with the atoms in C-layer, which characterizes the $\beta$-configuration. Ti-Al antisite defects in those structures induce partial transformation of $\alpha$ to $\beta$. Ti, Al and C atoms are denoted by green, red and yellow balls, respectively.

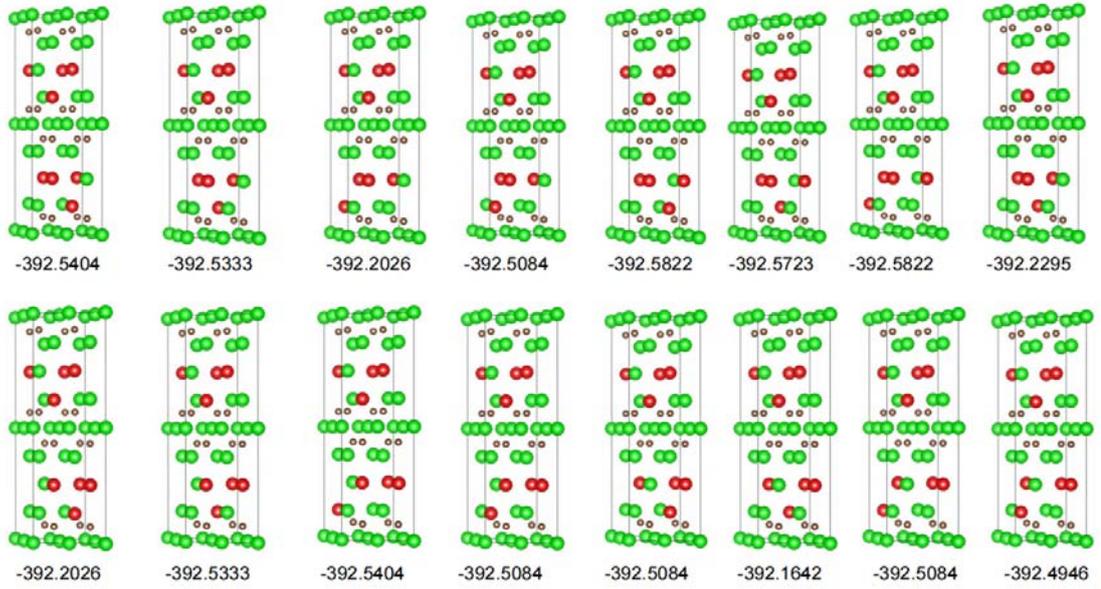

Fig. S11. Relaxed structures of β-Ti$_3$AlC$_2$ with 25% Al atoms forming Ti-Al antisite defects. The total energies in eV are denoted at the bottom. Ti, Al and C atoms are denoted by green, red and yellow balls, respectively.

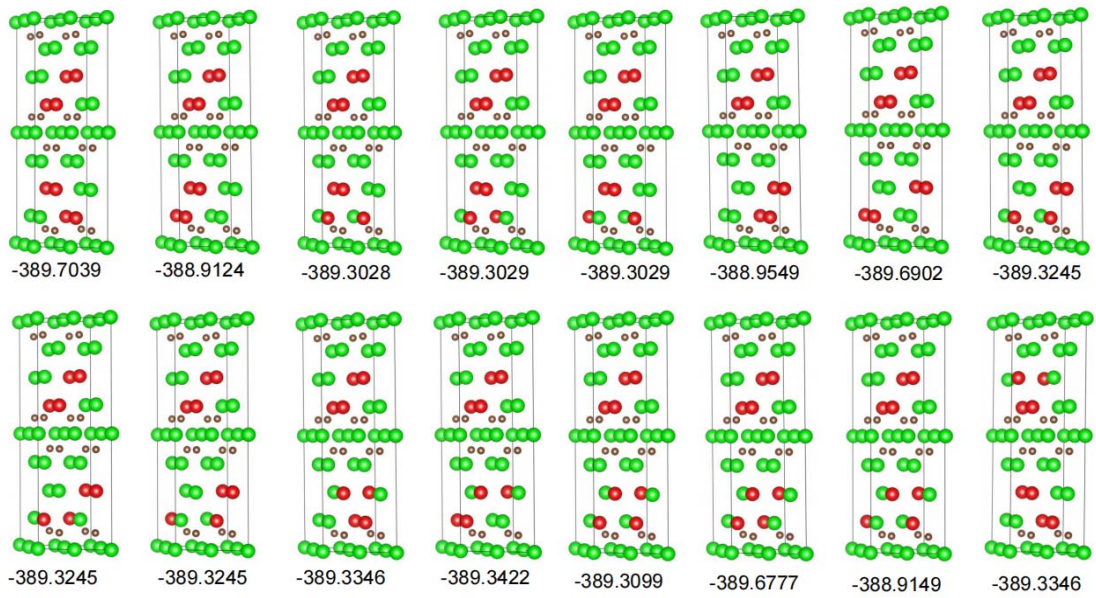

Fig. S12. Relaxed structures of β-Ti$_3$AlC$_2$ with 50% Al atoms forming Ti-Al antisite defects. The total energies in eV are denoted at the bottom. Ti, Al and C atoms are denoted by green, red and yellow balls, respectively.

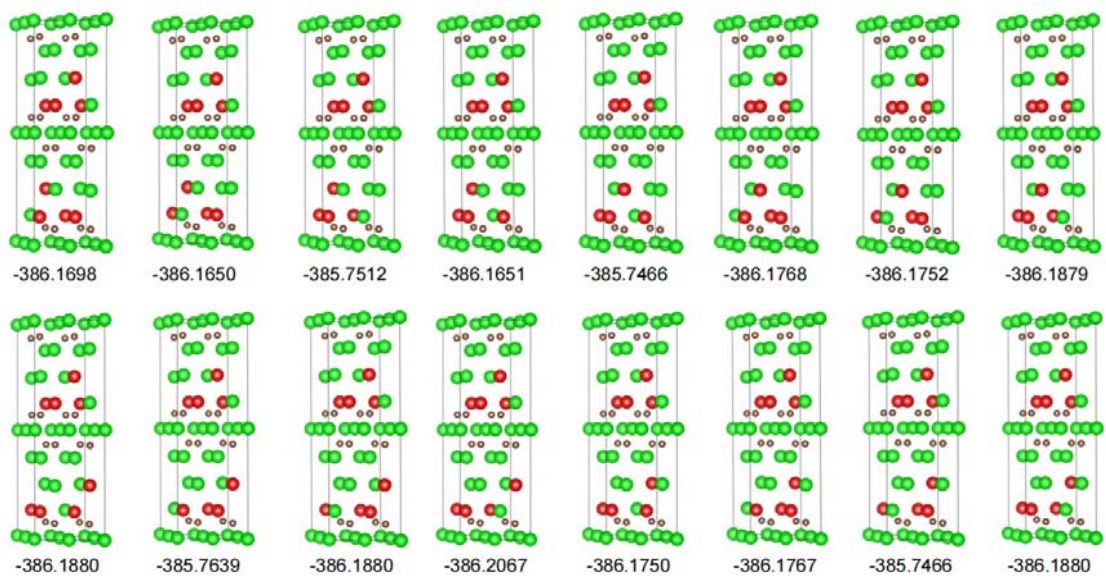

Fig. S13. Relaxed structures of β-Ti$_3$AlC$_2$ with 75% Al atoms forming Ti-Al antisite defects. The total energies in eV are denoted at the bottom. Ti, Al and C atoms are denoted by green, red and yellow balls, respectively.

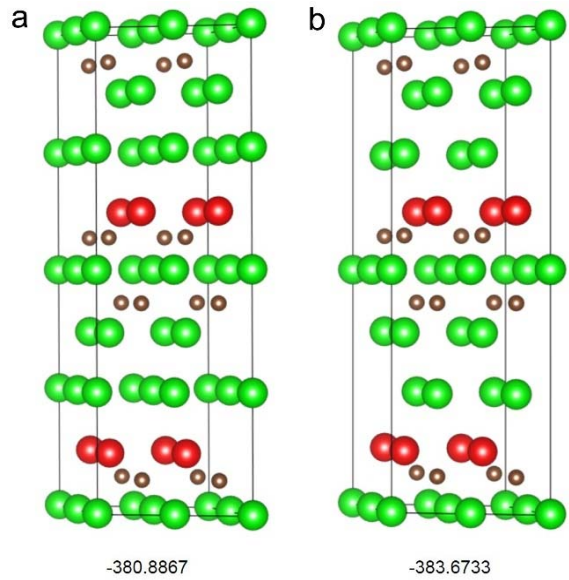

Fig. S14. Model used for electronic structure calculation in Fig. 4b,c. (a) $\alpha$- and (b) $\beta$-Ti$_3$AlC$_2$ with Al atoms in Al-layers completely replaced by Ti atoms just below Al-layers. The total energies in eV are denoted at the bottom. Ti, Al and C atoms are denoted by green, red and yellow balls, respectively.

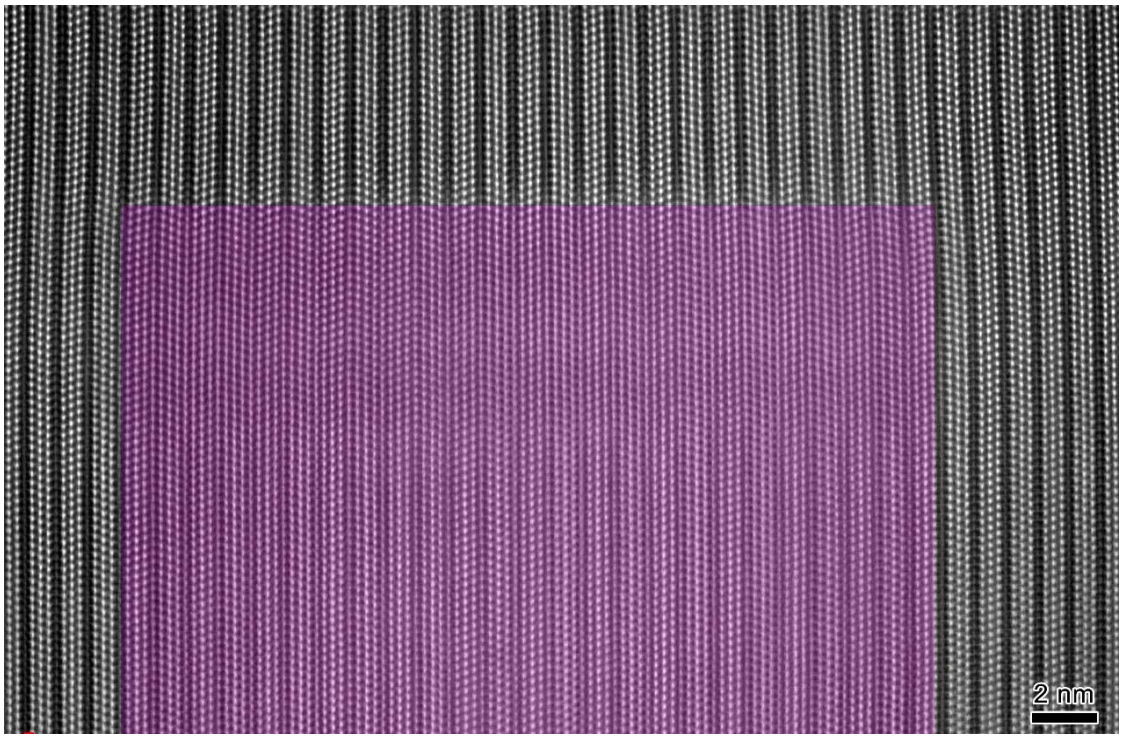

Fig. 15. Dilation after irradiation. The irradiated region (in pink) is clearly expanded along the *c*-axis (horizontal direction). From the fast Fourier transform, we can obtain lattice change. The structure dilates ~3.0% along the *c*-axis and contracts by 0.6% along the *a*-axis.

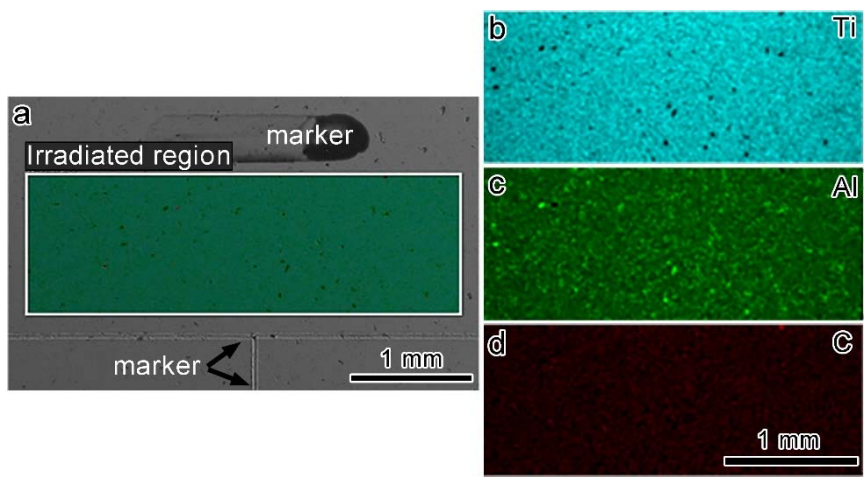

Fig. S16. Morphology and EDX of the irradiated sample. (a) Scanning electron microscopy image of electron irradiated sample. The shadowed rectangular (1.2 mm × 3.5 mm) region was irradiated. Markers were used for location identification. (b,c,d) EDX mapping of Ti, Al and C. Ti, Al and C are homogeneous except for some small Al-rich and Ti-poor regions, which might be $Al_2Ti$ or $Al_3Ti$ impurities in the as-prepared sample[1]. No TiC particles were observed. Large-area irradiation was conducted on SAILONG-S1 (Sailong Metal company, China) with an energy density of 14.4 J/cm$^2$. The chamber was vacuumed to 5 × 10$^{-3}$ Pa and then backfilled with helium gas to 0.15 Pa before irradiation.

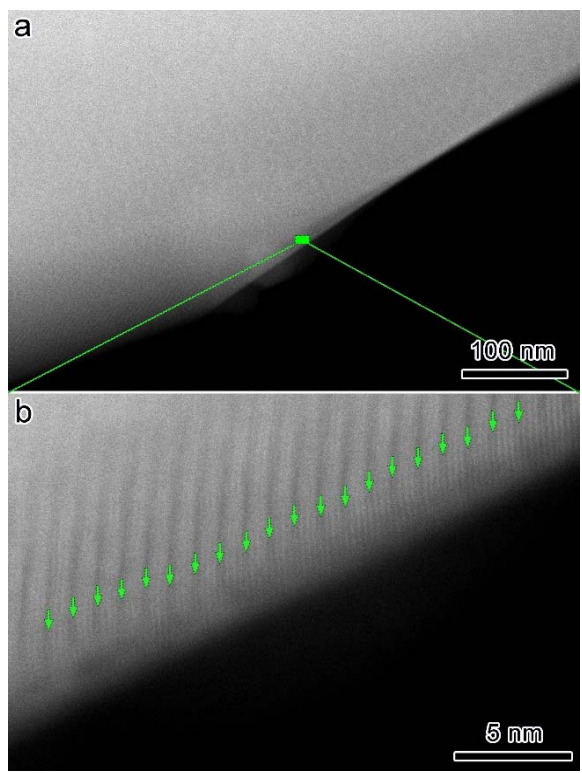

Fig. S17. HAADF-STEM images of cross-section sample. To observe the irradiated surface regions, the irradiated samples were prepared following the standard cross-section TEM sample preparation procedure. (a) is the low-magnification morphology. The high-magnification image in (b) was acquired with a dose of ~$10^5$ e$^-$/Å$^2$ to avoid extra electron irradiation as much as possible. Green arrows mark Al-layers. Layers between Al-layers are Ti-layers. Al-layers in the ~3 nm region near the surface show an intensity as strong as Ti-layers, indicating that Ti and Al were repartitioned, and nearly completely homogenized.

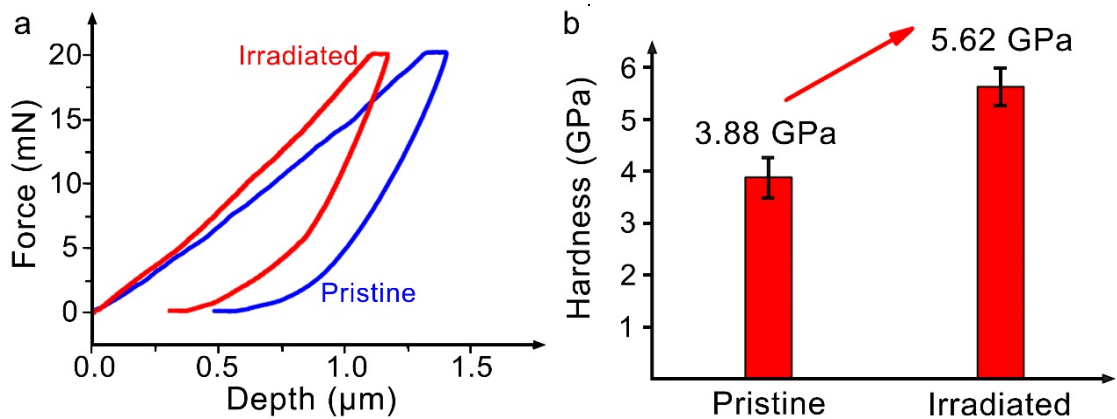

Fig. S18. Microhardness of the irradiated $Ti_3AlC_2$. (a) Typic load-displacement curves and (b) microhardness of the pristine and irradiated regions. The hardness before and after irradiation was measured on a dynamic ultra-micro hardness tester. The maximum load, loading rate and hold time were 20 mN, 0.8883 mN per second and five seconds, respectively. Five spots in the pristine and irradiated regions were randomly chosen for hardness measurement.

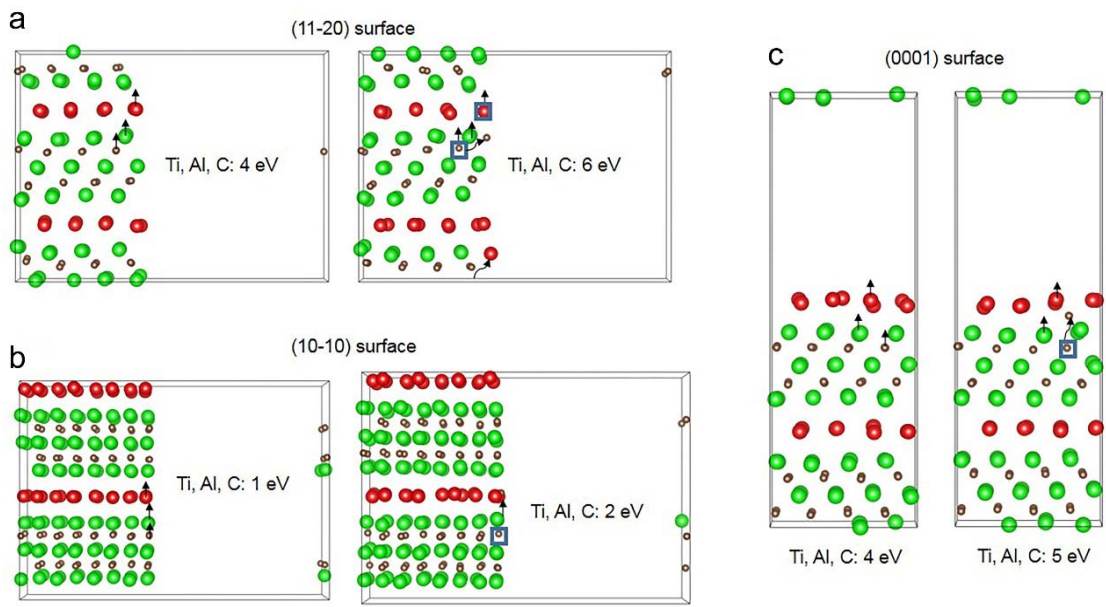

Fig. S19. Slab models used for the calculation of sputtering energies. Representative snapshots of $\alpha$-Ti$_3$AlC$_2$ (a) (11$\bar{2}$0), (b) (10$\bar{1}$0) and (c) (0001) with the atoms on the surface nonsputtered (left) and sputtered (right).

Table S1. Formation energies ($E_f$) of M/A antisite defects in MAX phases.

| Materials | $E_f$ (eV) | Other reports |
|---|---|---|
| $Ti_3AlC_2$ | 2.88 | 2.0[2], 3.13[3] |
| $Ti_2AlC$ | 2.71 | 2.96[3] |
| $Nb_2AlC$ | 2.49 | ~ |
| $Nb_4AlC_3$ | 3.27 | ~ |
| $Nb_2SC$ | 8.58 | ~ |

Table S2. Displacement threshold energies ($E_d$). $Ti_{Al}$, $Al_{Ti}$ and $C_i$ stand for Ti at Al sites, Al at Ti sites and interstitial C atoms, respectively.

| KPA | Direction | $E_d$ (eV) | Defects |
|---|---|---|---|
| Ti | [0001] | 37 | $Ti_{Al}$, $Al_{Ti}$ |
| Al | [0001] | 29 | $Ti_{Al}$, $Al_{Ti}$ |
| C | [0001] | 5 | $C_i$ |
| Ti, Al, C | [0001] | 13 | $Ti_{Al}$, $Al_{Ti}$, $C_i$ |

Table S3. Sputtering energies ($E_s$) of atoms on different surfaces. $C_i$ and $Al_i$ stand for interstitial C and Al, respectively.

| Surface plane | KPA | Direction | $E_s$ (eV) | Final defects |
|---|---|---|---|---|
| (11$\bar{2}$0) | Ti, Al, C | [0001] | 6 | $C_i$, $Al_i$ |
| (10$\bar{1}$0) | Ti, Al, C | [0001] | 2 | $C_i$ |
| (0001) | Ti, Al, C | [0001] | 5 | $C_i$ |

Table S4. The integrated crystal orbital hamilton populations (ICOHP) up to the Fermi level of aTi-sTi, aTi-aAl, aAl-C bonding for $\alpha$- and $\beta$-Ti$_3$AlC$_2$ with 100% Al replaced by Ti. The more negative ICOHP indicates a stronger interaction.

| ICOHP | $\alpha$-Ti$_3$AlC$_2$ | $\beta$-Ti$_3$AlC$_2$ |
|---|---|---|
| aTi-sTi | −0.04 | −1.31 |
| aTi-aAl | −0.03 | −1.21 |
| aAl-C | −4.34 | −4.48 |

Video S1. Structural evolution of $Ti_3AlC_2$ along $[1\bar{2}10]$ at 300 kV.

Video S2. Structural evolution of $Ti_3AlC_2$ along $[10\bar{1}0]$ at 300 kV.

Video S3. Structural evolution of $Ti_2AlC$ along $[1\bar{2}10]$ at 300 kV.

Video S4. Structural evolution of $Ti_3AlC_2$ and $Ti_2AlC$ along $[1\bar{2}10]$ at 300 kV.

Video S5: Structural evolution of $Ti_3AlC_2$ with one Ti, one Al and one C atoms added 13 eV along [0001]. The displaced Al first migrates into Ti-layer and pushes one Ti into Al-layer, forming $Al_{Ti}$ and $Ti_{Al}$ antisite pair. Whereas the displaced Ti moves back to its equilibrium position. The displaced C wanders around Al-layer and moves from one to another interstitial site. The $Al_{Ti}$ and $Ti_{Al}$ antisite pair forms at ~400 fs, and the interstitial C settles down at ~1300 fs. No recombination was observed.